\documentclass[runningheads]{llncs}
\usepackage[T1]{fontenc}
\usepackage{graphicx,verbatim}
\usepackage{hyperref}
\usepackage{booktabs}
\usepackage{multirow}
\usepackage{pifont}
\usepackage{graphicx}
\usepackage[table,xcdraw]{xcolor}
\usepackage{amsmath}
\usepackage{amssymb}

\begin{document}

\title{XTinyU-Net: Training-Free U-Net Scaling via Initialization-Time Sensitivity}
%


\author{Alvin Kimbowa\inst{1}\textsuperscript{*}
\and
Moein Heidari\inst{1}\textsuperscript{*}
\and
David Liu\inst{1,2}
\and
Ilker Hacihaliloglu\inst{1,2,3}
}
\authorrunning{A. Kimbowa et al.}
%
\institute{School of Biomedical Engineering, The University of British Columbia, Vancouver, Canada \and
Department of Radiology, The University of British Columbia, Vancouver, Canada \and
Department of Medicine, The University of British Columbia, Vancouver, Canada\\
\email{alvinbk@student.ubc.ca}\\
\email{\{moein.heidari, david.liu, ilker.hacihaliloglu\}@ubc.ca}\\
}

\maketitle              
\begingroup
\renewcommand{\thefootnote}{\fnsymbol{footnote}}
\footnotetext[1]{Equal contribution}
\endgroup

\begin{abstract}
While U-Net architectures remain the gold standard for medical image segmentation, their deployment in resource-constrained environments demands aggressive model compression. However, finding an optimally efficient configuration is computationally prohibitive, typically requiring exhaustive train-and-evaluate cycles to find the smallest model that maintains peak performance. In this paper, we introduce a training-free selection framework to automatically identify ultra-lightweight, dataset-specific U-Net configurations directly at initialization. We observe that systematically scaling down U-Net channel width induces a sharp transition from a stable performance plateau to representational capacity collapse. To pinpoint this boundary without training, we propose a Jacobian-based sensitivity metric that scores discrete, width-capped U-Net variants using a small set of unlabeled images. By analyzing the total variation of this sensitivity curve, we isolate the smallest stable configuration, which we denote as XTinyU-Net. Evaluated across six diverse medical datasets within the nnU-Net framework, XTinyU-Net achieves segmentation accuracy comparable to the heavy nnU-Net baseline with 400$\times$–1600$\times$ fewer parameters, and outperforms contemporary lightweight architectures while utilizing 5$\times$–72$\times$ fewer parameters. Code is publicly accessible on \href{https://github.com/alvinkimbowa/nntinyunet.git}{GitHub}.


\end{abstract}

\section{Introduction}
Medical image segmentation is a foundational primitive for quantitative clinical analysis across diverse modalities~\cite{azad2024medical,moor2023foundation}. While modern deep neural networks achieve remarkable fully-supervised performance, the deployment landscape is rapidly shifting toward decentralized, resource-constrained environments such as point-of-care systems and portable devices. To bridge the gap between high accuracy and low computational cost, numerous lightweight architectures have been proposed. These specialized models reduce parameters and floating-point operations via compact blocks, simplified decoders, or token mixing, as exemplified by UNeXt~\cite{Valanarasu2022UNeXt}, CMUNeXt~\cite{Tang2024CMUNEXT}, MedNCA~\cite{Kalkhof2025MEDNCA}, TinyUNet~\cite{Che_TinyUNet_MICCAI2024}, and LeanUNet~\cite{Hassler2025LeanUnet}. Despite this architectural proliferation, the canonical U-Net~\cite{RonnebergerFB15} remains an exceptionally competitive baseline, frequently outperforming specially crafted lightweight models when optimized with standardized pipelines like nnU-Net~\cite{Isensee2021nnunet}.

These lightweight methods demonstrate that strong accuracy can be achieved at low cost, yet they surface an inconvenient reality: efficiency is not a static, one-size-fits-all architectural choice. For a fixed model family, the optimal budget depends heavily on dataset-specific characteristics, including image resolution, anatomical variability, noise, and label complexity. Consequently, practitioners still rely on computationally prohibitive manual sweeps to train and evaluate multiple candidate configurations. This expensive validation process undermines the core promise of lightweight models in settings where rapid iteration is essential.

This paper targets the configuration selection problem directly, asking: \textit{How aggressively can a properly configured U-Net be scaled down for a specific dataset before representational capacity collapses and segmentation performance degrades?} We focus on width scaling because channel dimensions dominate computation in U-Net-style models, and it serves as the primary knob in many lightweight designs. To isolate the effect of capacity without confounding factors, we instantiate our study within the standardized nnU-Net protocol.

To address this selection bottleneck without training, our approach connects to zero-cost neural architecture search (NAS)~\cite{pmlr-v139-mellor21a,chen2020tenas,Lin_2021_ICCV}. Existing zero-cost methods predict downstream accuracy using initialization-time statistics, with representative directions including activation-based criteria (e.g., NASWOT~\cite{pmlr-v139-mellor21a}, SWAP~\cite{Peng2024SWAPNAS}) and gradient-based criteria (e.g., SynFlow~\cite{Tanaka2020SynFlow}, SNIP~\cite{Lee2019SNIP}). However, these influential approaches are typically evaluated on classification tasks and macro-level inter-architecture comparisons. In contrast, our intra-family, dense-prediction setting requires reliably evaluating small architectural changes where capacity regime transitions are highly pronounced. Rather than treating the score as a global ranker, we utilize the Jacobian signal as a targeted collapse detector to answer the practical deployment question: how small can the model be before performance deteriorates?

Specifically, our contributions are as follows: \ding{202} We systematically analyze aggressive U-Net capacity scaling and consistently observe a predictable two-regime behavior: a broad performance plateau followed by an abrupt representational collapse at ultra-small model sizes. \ding{203} Building on this observation, we introduce a training-free, Jacobian-based framework for selecting dataset-adaptive lightweight configurations directly at initialization. This robust metric instantly isolates the optimal pre-collapse transition configuration, which we designate as XTinyU-Net. \ding{204} Extensive evaluations across six diverse medical imaging datasets establish that XTinyU-Net achieves segmentation accuracy comparable to the nnU-Net baseline while requiring up to 1600x fewer parameters, and it consistently outperforms specialized state-of-the-art lightweight architectures while utilizing 5x–72x fewer parameters.

\section{Method}
We frame the search for an optimal lightweight segmentation network as a problem of detecting capacity collapse within a progressively scaled architecture family. Rather than relying on expensive, iterative train-and-evaluate cycles , we propose a training-free selection mechanism that operates entirely at random initialization. Our approach proceeds in two main steps. First, we construct a discrete, ordered family of candidate architectures by systematically constraining the maximum channel width of a standard baseline. Second, we evaluate the input-output sensitivity of each untrained candidate using a Jacobian-based metric. By analyzing the total variation of this initialization-time sensitivity across the model family, we can pinpoint the critical capacity boundary where the network transitions from a stable performance plateau to an unstable, post-collapse regime. The configuration lying immediately on the stable side of this boundary is selected as our highly efficient target architecture, XTinyU-Net

\subsection{U-Net scaling}
\label{sect:unet_scaling}

We systematically investigate U-Net capacity limits by scaling network width (channels) rather than depth (stages).
Reducing depth alters the model's inductive bias by shrinking the theoretical receptive field and disrupting multi-scale context aggregation.
In contrast, width scaling provides a strictly controlled axis for compression: it isolates representational capacity while perfectly preserving the architecture's foundational spatial topology and downsampling schedule.

Let $C_l^{\text{base}} = 2^l C_0$, where $C_0$ is the number of channels at the first stage and $L$ is the total number of stages, and let $C_{\max}^{\text{base}} = C_{L-1}^{\text{base}}$ denote the maximum channel dimensionality in the baseline configuration.
We define a discrete family of scaled models by imposing a geometric channel cap
\begin{equation}
C_{\max}(i) = \frac{C_{\max}^{\text{base}}}{2^i},i \in \{0,1,\dots,\log_2 C_{\max}^{\text{base}}\}
\end{equation}
such that the channel width at stage $l$ is then given by
$C_l(i) = \min\!\left(2^l C_0,\; C_{\max}(i)\right)$.

As $i$ increases, the maximum allowable channel dimensionality decreases, yielding progressively smaller models.
For large values of $i$, the doubling pattern across resolutions saturates and the network approaches a near-uniform channel dimensionality across stages as in~\cite{Hassler2025LeanUnet}. Importantly, the spatial hierarchy, number of stages, and receptive-field growth remain unchanged, and all other architectural and optimization settings are kept fixed.
We use $C_{max}^{base}=512$ as in nnU-Net~\cite{Isensee2021nnunet} and thus end up with a family of 10 models.
We denote the resulting ordered model family by 
$\{\mathcal{M}_i\}_{i=0}^{N-1}$, 
where increasing index $i$ corresponds to progressively smaller channel caps and reduced model capacity.

\begin{figure}[t]
    \centering
    \includegraphics[width=\linewidth]{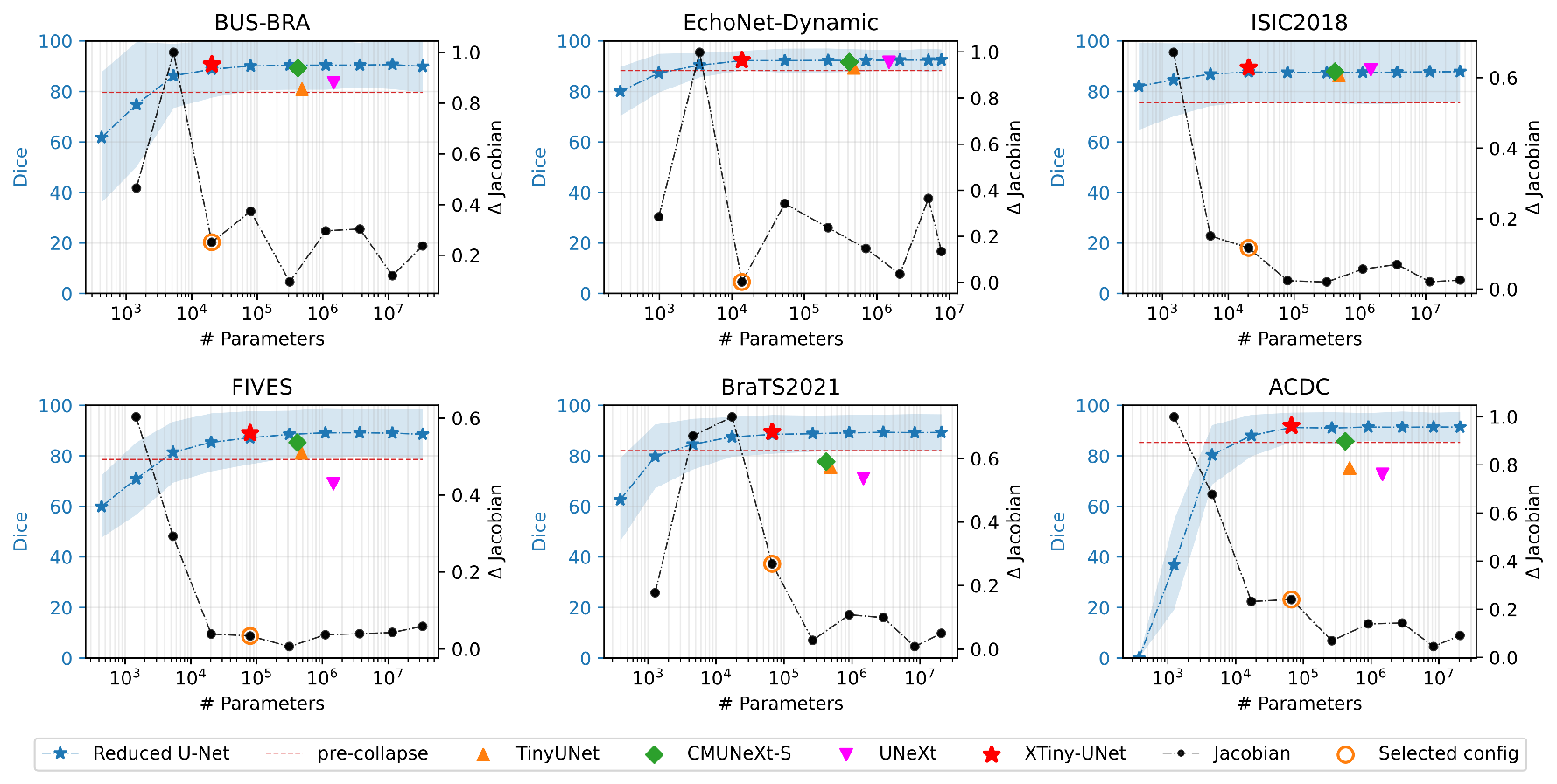}
    \caption{U-Net width scaling curves (Dice vs parameters) and corresponding initialization-time Jacobian sensitivity across datasets. The estimated collapse boundary separates stable and unstable regimes; XTinyU-Net is selected as the smallest configuration on the stable side.}
    \label{fig:main}
\end{figure}

When training the defined U-Net family across datasets, we consistently observe two distinct performance regimes (measured by Dice, as seen in Fig. \ref{fig:main}) as model size decreases.
For moderate width reductions, segmentation performance remains relatively stable, forming a broad plateau. Beyond a dataset-dependent threshold, performance deteriorates sharply as models become severely underparameterized. We refer to these regimes as the pre-collapse and post-collapse regions, respectively.
Different datasets exhibit different transition points, reflecting varying task complexity and representational requirements.
This naturally raises the following question:
\textit{"Given a medical imaging dataset, can we estimate the smallest configuration that remains within the pre-collapse regime, at initialization without training the entire model family?"}

\subsection{Jacobian-Based Collapse Detection}
Our hypothesis is that the transition from the pre-collapse to the post-collapse regime is reflected in the local input--output sensitivity of the network at initialization. As model capacity decreases, sufficiently narrow configurations exhibit increased sensitivity to small perturbations in the input, indicating unstable representational behavior near the capacity boundary.
\subsubsection{Input--output sensitivity.}
To quantify how sensitive a given model $M_i$ (from the model family defined in Section \ref{sect:unet_scaling}), is to a given dataset at initialization, we sample $K$ images from the dataset, compute the gradient of the model output with respect to each input image, followed by a computation of a scalar sensitivity score that serves as the sensitivity estimate for the entire dataset.
Specifically, for the sampled input images $x\in\mathbb{R}^{K\times C\times H \times W}$, where $C$ is the number of channels, and $H,W$ are the height and width, respectively, we take the gradient of the model output as $G_i(x) = \nabla_{x} \sum_{u} M_i(x;\theta_i)_u$,
where $M_i(x;\theta_i)\in\mathbb{R}^{K\times N\times H \times W}$, $N$ is the number of output classes, $u$ indexes all output spatial locations and class channels, and $G_i(x)\in \mathbb{R}^{K\times C \times H\times W}$.
We then define a scalar sensitivity measure across the sampled images using a root-mean-square (RMS) estimate
$S_i(x) =
\sqrt{
\frac{1}{K}
\sum_{k=1}^{K}
\left\|G_i^k(x)\right\|_2^2
}.$

\subsubsection{Cross-configuration normalization.}
Since the sensitivity magnitudes are dimensionless and arbitrary, with $S_i\in[0,\infty)$, we first normalize the values across the model configurations for meaningful comparison as follows; $\tilde{S}_i = \{{S_i - S_{\min}\}}/\{{S_{\max}- S_{\min}\}}$. Empirically, we observe that in the unstable post-collapse regime, the changes in the input-output sensitivity values between consecutive model configurations are larger compared to the changes in the pre-collapse region.

\subsubsection{Collapse detection.}
Thus, we define the transition configuration between the pre- and post-collapse regimes to be the one that maximizes the total variation in the Jacobians in the post-collapse regime while minimizing the total variation in the pre-collapse regime.
Let $\{M_i\}_{i=0}^{N-1}$ be model configurations ordered from largest capacity (i=0) to smallest capacity (i=N-1),
and let $S_i$ denote the sensitivity score of $M_i$.
We define the local variation between consecutive model configurations as$d_i = |S_{i+1} - S_i|$, where $i \in \{1,N-1\}$.
For a candidate split index $k \in \{2,\dots,N-1\}$, we define the total variation on the post-collapse (small models) and pre-collapse (large models) regions as
\[
\mathrm{TV}_L(k) = \sum_{i=0}^{k-1} d_i,
\qquad
\mathrm{TV}_R(k) = \sum_{i=k}^{N-1} d_i,
\]
respectively.
We select the collapse boundary as
\[
k^\star = \arg\max_{k} \bigl( \mathrm{TV}_R(k) - \mathrm{TV}_L(k) \bigr),
\]
and choose $M_{k^\star}$ as the final pre-collapse configuration of choice.
We dub this configuration as \textbf{XTinyU-Net}.
This way, we obtain a lightweight architecture that is as close to the transition point from pre- to post-collapse as possible (corresponding to red stars in Fig.~\ref{fig:main}).

\section{Experiments}

\paragraph{\textbf{{Datasets}}}
\label{sec:datasets}
We evaluate our method on six public medical image segmentation datasets spanning diverse modalities. Specifically, we utilize BUS-BRA~\cite{gomez2024bus}, EchoNet Dynamic~\cite{ouyang2020video}, and ISIC 2018~\cite{codella2019skin}.
We resize the ISIC data to $256 \times 256$ to reduce its resolution due to computational constraints.
Furthermore, we include FiVES~\cite{jin2022fives} for retinal fundus vessels, ACDC~\cite{bernard2018deep} for cardiac cine MRI structures, and BraTS~\cite{menze2014multimodal} for brain MRI tumor sub-regions.
We maintain the default splits of the datasets to ensure fairness across evaluations.

\paragraph{\textbf{{Implementation details}}}
We instantiate our approach within the nnU-Net training and evaluation protocol in order to keep preprocessing, augmentation, patch size, and optimization fixed across all configurations.
This ensures that performance differences primarily reflect capacity changes rather than variations in the training recipe.
All models are trained using the default nnU-Net settings for each dataset. We report Dice and the 95th percentile Hausdorff distance (HD95), and present results as mean and standard deviation over three runs.
We compare against recent efficient segmentation architectures targeting low-compute settings, including UNeXt, CMUNeXt-S, and TinyU-Net. When available, nnU-Net is included as a reference under the same protocol.

\section{Results and Discussion}

\begin{table*}[t]
\centering
\caption{Results on BUS-BRA, EchoNet Dynamic, and ISIC 2018. We report Params, FLOPs, Dice, and HD95 (mean$\pm$std over three runs). Best is bold and second best is underlined. nnU-Net is shown in gray as the upper bound.}
\renewcommand{\arraystretch}{1.1}
\resizebox{\textwidth}{!}{%
\begin{tabular}{@{}lcccccccc@{}}
\toprule
 &  &  & \multicolumn{2}{c}{\textbf{BUS-BRA}} & \multicolumn{2}{c}{\textbf{EchoNet Dynamic}} & \multicolumn{2}{c}{\textbf{ISIC2018}} \\ \cmidrule(l){4-9} 
\multirow{-2}{*}{\textbf{Config.}} & \multirow{-2}{*}{\textbf{\begin{tabular}[c]{@{}c@{}}Params\\ (k)$\downarrow$\end{tabular}}} & \multirow{-2}{*}{\textbf{\begin{tabular}[c]{@{}c@{}}FLOPS\\ (G)$\downarrow$\end{tabular}}} & \textbf{Dice}$\uparrow$ & \textbf{HD95}$\downarrow$ & \textbf{Dice}$\uparrow$ & \textbf{HD95}$\downarrow$ & \textbf{Dice}$\uparrow$ & \textbf{HD95}$\downarrow$ \\ \midrule
UNeXt~\cite{Valanarasu2022UNeXt} & 1471.94 & 4.59 & 83.43±13.78 & 20.19±20.77 & \underline{91.52±3.91} & 12.09±13.99 & 88.58±12.86 & \underline{20.64±23.33} \\
CMUNeXt-S~\cite{Tang2024CMUNEXT} & \underline{417.50} & \underline{2.18} & \underline{89.22±10.32} & \textbf{12.47±15.89} & 90.81±5.54 & \underline{11.65±12.08} & \underline{88.67±13.53} & 21.02±26.20 \\
TinyU-Net~\cite{Che_TinyUNet_MICCAI2024} & 481.11 & 3.32 & 80.88±17.91 & 27.97±27.30 & 89.57±5.74 & 18.35±18.30 & 86.37±15.48 & 24.25±26.35 \\ \midrule
\textbf{XTinyU-Net} & \textbf{20.39} & \textbf{0.92} & \textbf{90.64±7.57} & \underline{14.74±20.43} & \textbf{92.31±3.94} & \textbf{3.78±2.52} & \textbf{89.39±12.38} & \textbf{18.19±23.56} \\ \midrule
{\color[HTML]{9B9B9B} nnU-Net~\cite{Isensee2021nnunet}} & {\color[HTML]{9B9B9B} 33,472.40} & {\color[HTML]{9B9B9B} 62.40} & {\color[HTML]{9B9B9B} 89.90±10.25} & {\color[HTML]{9B9B9B} 15.25±21.05} & {\color[HTML]{9B9B9B} 92.45±4.16} & {\color[HTML]{9B9B9B} 3.75±3.90} & {\color[HTML]{9B9B9B} 87.84±12.19} & {\color[HTML]{9B9B9B} 19.63±23.32} \\ \bottomrule
\end{tabular}
}
\label{tab:easy}
\end{table*}

\begin{table*}[t]
\centering
\caption{Results on FiVES, BraTS2020, and ACDC. We report Params, FLOPs, Dice, and HD95 (mean$\pm$std). Best is bold and second best is underlined. 
}
\setlength{\tabcolsep}{3.2pt}
\renewcommand{\arraystretch}{1.1}
\resizebox{\textwidth}{!}{%
\begin{tabular}{@{}lcccccccc@{}}
\toprule
 &  &  & \multicolumn{2}{c}{\textbf{FIVES}} & \multicolumn{2}{c}{\textbf{BraTS2020}} & \multicolumn{2}{c}{\textbf{ACDC}} \\ \cmidrule(l){4-9} 
\multirow{-2}{*}{\textbf{Config.}} & \multirow{-2}{*}{\textbf{\begin{tabular}[c]{@{}c@{}}Params\\ (k)$\downarrow$\end{tabular}}} & \multirow{-2}{*}{\textbf{\begin{tabular}[c]{@{}c@{}}FLOPS\\ (G)$\downarrow$\end{tabular}}} & \textbf{Dice}$\uparrow$ & \textbf{HD95}$\downarrow$ & \textbf{Dice}$\uparrow$ & \textbf{HD95}$\downarrow$ & \textbf{Dice}$\uparrow$ & \textbf{HD95}$\downarrow$ \\ \midrule
UNeXt~\cite{Tang2024CMUNEXT} & 1471.67 & 4.59 & 68.98±7.14 & 13.25±10.71 & 71.08±17.81 & 26.71±16.26 & 72.00±14.28 & 8.38±6.52 \\
CMUNeXt-S~\cite{Tang2024CMUNEXT} & \underline{417.60} & \underline{2.19} & \underline{85.27±9.80} & \underline{5.99±12.97} & \underline{77.68±37.43} & \underline{12.33±12.54} & \underline{84.40±20.58} & \underline{4.06±4.91} \\
TinyU-Net~\cite{Che_TinyUNet_MICCAI2024} & 481.31 & 3.34 & 84.72±8.07 & 6.99±12.63 & 75.43±17.75 & 17.48±20.02 & 71.28±22.29 & 14.35±17.19 \\ \midrule
\textbf{XTinyU-Net} & \textbf{79.55} & \textbf{1.96} & \textbf{88.97±10.42} & \textbf{5.17±14.80} & \textbf{89.47 ± 11.30} & \textbf{4.15±5.27} & \textbf{91.86±6.06} & \textbf{2.53±8.98} \\ \bottomrule
{\color[HTML]{9B9B9B} nnU-Net~\cite{Isensee2021nnunet}} & {\color[HTML]{9B9B9B} 33,472.40} & {\color[HTML]{9B9B9B} 33.36} & {\color[HTML]{9B9B9B} 88.57±10.03} & {\color[HTML]{9B9B9B} 5.08±13.13} & {\color[HTML]{9B9B9B} 91.19±11.17} & {\color[HTML]{9B9B9B} 2.89±3.33} & {\color[HTML]{9B9B9B} 91.37±6.01} & {\color[HTML]{9B9B9B} 2.27±6.65} \\ \bottomrule
\end{tabular}
}
\label{tab:hard}
\end{table*}

\subsection{Quantitative Results}

\subsubsection{U-Net scaling.}
Across all six datasets, width scaling reveals a consistent two-regime pattern. As model capacity decreases, Dice remains relatively stable over a broad range of channel caps before degrading sharply at very small widths. This data-dependent pattern is visible in Fig.~\ref{fig:main}.

\subsubsection{XTinyU-Net.}
XTinyU-Net consistently selects a configuration near the empirical collapse boundary while remaining on the stable side of the transition.
As shown in Tables~\ref{tab:easy} and ~\ref{tab:hard}, XTinyU-Net achieves performance comparable to nnU-Net across datasets despite using over 400x–1600x fewer parameters.
For instance, on BUS-BRA, XTinyU-Net achieves 90.64\% Dice compared to 89.90\% for nnU-Net while reducing parameters from 33,472k to 20.39k (over $1600\times$ fewer parameters) and FLOPs from 62.40G to 0.92G ($\sim68\times$ reduction). Similar trends are observed on EchoNet-Dynamic and ACDC, where Dice remains within a narrow margin of the baseline despite substantial computational savings.

\subsubsection{Baseline comparison.}
Compared to recent efficient segmentation models (UNeXt, CMUNeXt-S, TinyU-Net), XTinyU-Net achieves competitive or superior Dice while using over 5x-72x fewer parameters.
These results suggest that dataset-adaptive width selection can be as effective as architecture redesign for achieving favorable performance-efficiency trade-offs.

\begin{figure}[t]
    \centering
    \includegraphics[width=0.85\linewidth]{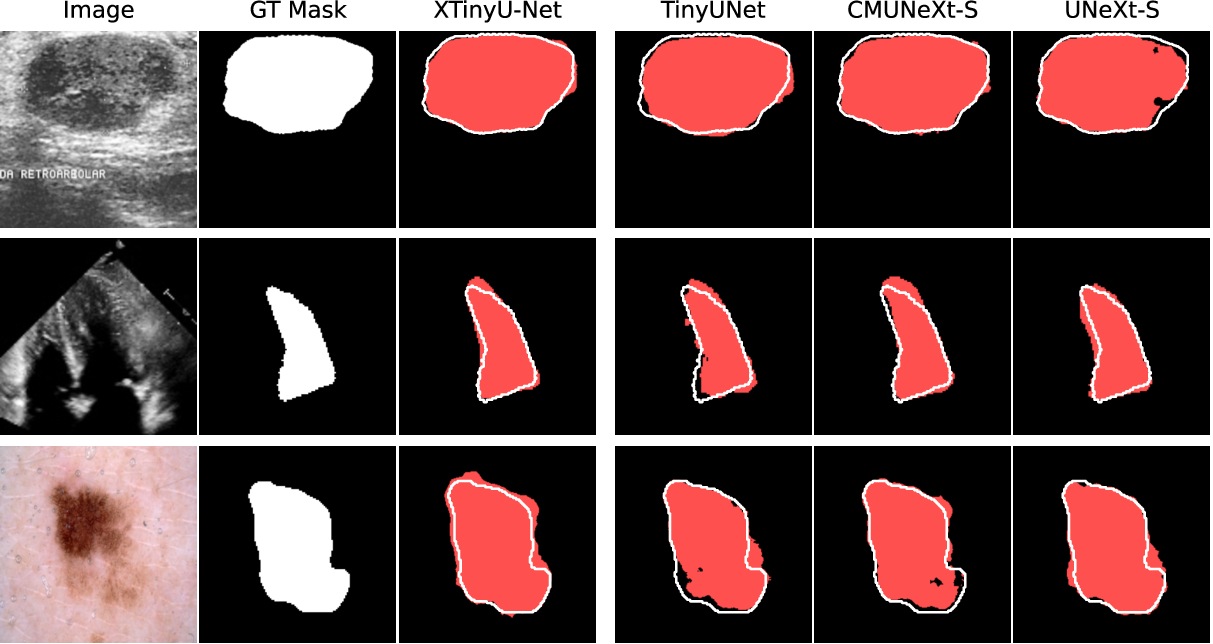}
    \caption{Qualitative results on BUS-BRA (top row), EchoNet-Dynamic (middle row), and ISIC 2018 (bottom row). XTinyU-Net achieves high quality segmentation performance.}
    \label{fig:easy}
\end{figure}

\begin{figure}[t]
    \centering
    \includegraphics[width=0.85\linewidth]{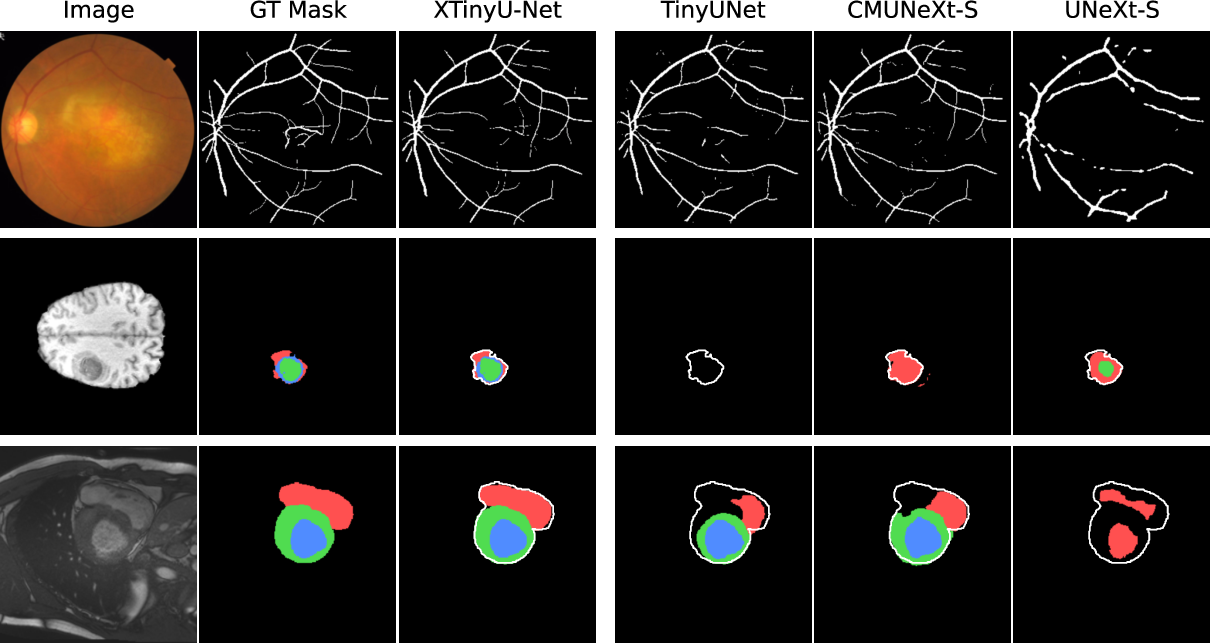}
    \caption{Qualitative results on FiVES, BraTS2020, and ACDC (bottom row). XTinyU-Net achieves high quality segmentation performance.
    }
    \label{fig:hard}
\end{figure}

\subsection{Qualitative Results}

Representative segmentation outputs are shown in Figs.~\ref{fig:easy} and~\ref{fig:hard}. Despite having significantly higher parameters, baseline light-weight architectures exhibit fragmented masks, missing regions, and unstable boundaries. In contrast, XTinyU-Net generates cleaner and more complete segmentations that closely resemble ground-truth masks.

\subsection{Ablation Study}

\begin{figure}
    \centering
    \includegraphics[width=\linewidth]{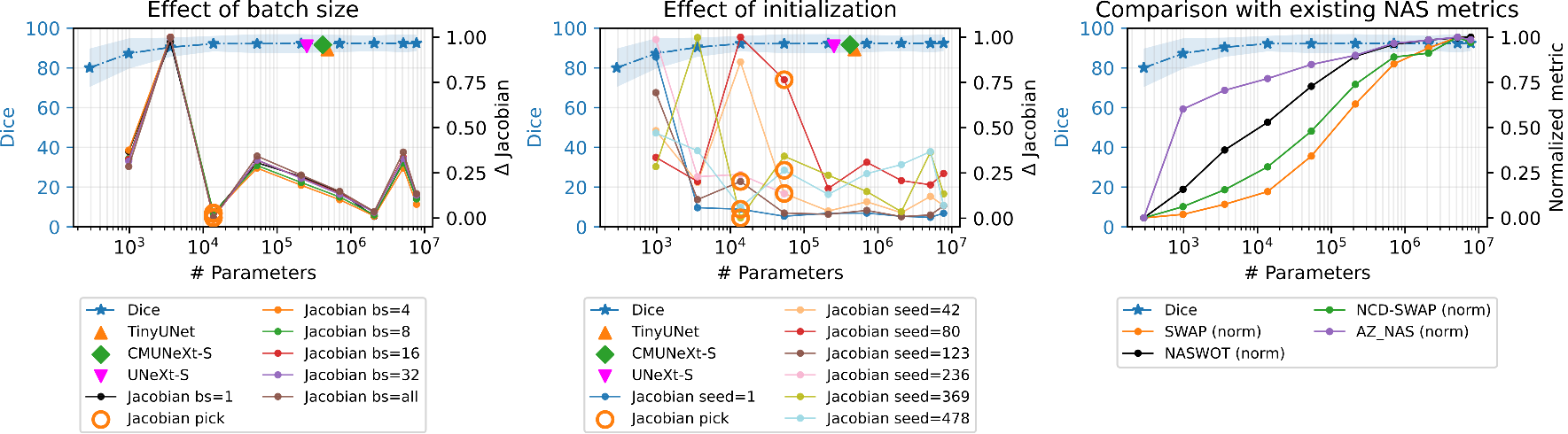}
    \caption{Ablation study on EchoNet-Dynamic showing robustness of the proposed Jacobian-based collapse detection method to batch size, random initialization, and comparison with existing NAS metrics.}
    \label{fig:ablation}
\end{figure}

As illustrated in Fig.~\ref{fig:ablation}, we evaluate the robustness of our Jacobian-based collapse detection and compare it against existing zero-cost NAS metrics. \textbf{Batch size \& Initialization:} Our method exhibits robustness to batch size variations, consistently identifying the same model configuration even with a single input image. It also demonstrates resilience to random initialization seeds, with the selected architectures tightly clustered around two adjacent configurations. Crucially, these selected models achieve competitive segmentation performance while requiring 5x to 72x fewer parameters than existing lightweight networks. \textbf{NAS Metric Comparison:} We observe that established NAS metrics (e.g., NASWOT, SWAP) scale monotonically with model capacity. This inherently continuous increase renders existing metrics incapable of detecting the sharp pre- to post-collapse transition boundary in U-Net architectures. Conversely, our sensitivity measure exhibits a distinct peak that successfully isolates this critical boundary.

\section{Conclusion}

We propose a training-free, Jacobian-based collapse detection method to identify optimal, dataset-specific U-Net configurations directly at initialization. Evaluated across six medical datasets, our resulting model, XTinyU-Net, matches the segmentation performance of nnU-Net while requiring up to 1600x fewer parameters. Furthermore, it outperforms specialized lightweight architectures using 5x to 72x fewer parameters. Ultimately, we demonstrate that a properly width-scaled standard U-Net provides a superior performance-efficiency trade-off compared to custom lightweight designs.

\section{Acknowledgments}
This work was supported by the Canadian Foundation for Innovation-John R. Evans Leaders Fund (CFI-JELF) program [Grant ID 42816, AWD-023869 CFI].
We acknowledge the support of the Natural Sciences and Engineering Research Council of Canada (NSERC), [RGPIN-2023-03575 AWD-024385]. Cette recherche a été financée par le Conseil de recherches en sciences naturelles et en génie du Canada (CRSNG), [RGPIN-2023-03575, AWD-024385].
We acknowledge the support provided by the Canadian Consortium of Clinical Trial Training (CANTRAIN) platform, Michael Smith Health Research British Columbia, and the Minimally Invasive Image Guided Procedure Lab (MIIPs), at the University of British Columbia School of Biomedical Engineering.

\bibliographystyle{splncs04}
\bibliography{refs}

\end{document}